\documentclass[preprint,12pt]{elsarticle}
\usepackage{graphicx,color,mathrsfs,amsmath,amssymb,amsthm,amsopn,bm}
\usepackage{epsfig}
\newcommand{\be}{\begin{equation}}
\newcommand{\ee}{\end{equation}}
\newcommand{\bge}{\begin{equation}}
\newcommand{\ene}{\end{equation}}
\newcommand{\bea}{\begin{eqnarray}}
\newcommand{\eea}{\end{eqnarray}}
\newcommand{\bg}{\begin{eqnarray}}
\newcommand{\en}{\end{eqnarray}}

\renewcommand{\vec}[1]{\boldsymbol{#1}}

\def\eqref#1{Eq.~(\ref{eq:#1})}

\def\be{\begin{equation}}
\def\ee{\end{equation}}
\def\bg{\begin{eqnarray}}
\def\en{\end{eqnarray}}
\def\nn{\nonumber}

\long\def\Omit#1{}

\journal{Physics Letters B}

\begin{document}

\begin{frontmatter}

\title{Production of $\Lambda_c^+$ hypernuclei in antiproton - nucleus collisions}
\author[Saha]{R.~Shyam}
\author[LFTC]{K.~Tsushima}
\address[Saha]{Saha Institute of Nuclear Physics, 1/AF Bidhan Nagar, Kolkata 700064, 
India}
\address[LFTC]{Laborat\'{o}rio de F\'{i}sica Te\'{o}rica e Computacional, Universidade 
Cruzeiro do Sul, Rua Galv\~{a}o Bueno, 868, Liberdade 01506-000, S\~{a}o Paulo, SP, 
Brazil } 

\date{\today}

\begin{abstract}

We investigate the production of charm-baryon hypernucleus $^{16}_{\Lambda_c^+}$O
in the antiproton - $^{16}$O collisions within a fully covariant model that
is based on an effective Lagrangian approach. The explicit ${\bar \Lambda}_c^
- \Lambda_c^+$ production vertex is described by the $t$-channel $D^0$ and 
$D^{*0}$ meson-exchanges in the initial collision of the incident antiproton with 
one of the protons of the target nucleus. The $\Lambda_c^+$ bound state spinors 
as well as the self-energies of the exchanged mesons employed in our calculations 
are derived from the quark-meson coupling model. The parameters of various 
vertices are taken to be the same as those used in our previous study of the 
elementary ${\bar p} + p \to {\bar \Lambda}_c^- + \Lambda_c^+$ reaction. We 
find that for antiproton beam momenta of interest to the ${\bar P}ANDA$ 
experiment, the 0$^\circ$ differential cross sections for the formation of 
$^{16}_{\Lambda_c^+}$O hypernuclear states with simple particle-hole 
configurations, have magnitudes in the range of a few $\mu$b/sr.

\end{abstract}


\end{frontmatter}

The investigation of the production of heavy flavor hadrons consisting of 
a charm-quark is of considerable interest as it provides an additional means
for a better understanding of quantum chromodynamics (QCD) (see, e.g., Refs.
\cite{neu94,kor94}). The future ${\bar P}ANDA$ ("antiproton annihilation at 
Darmstadt") experiment at the under-construction antiproton and ion research 
facility (FAIR) in Darmstadt, Germany, includes a rich program on the 
measurements of the charm-meson and charm-baryon production in the antiproton 
(${\bar p}$) collisions with proton and nuclei at the beam momenta $\leq$ 15 
GeV/c~\cite{pan09}. The accurate knowledge of the charm-meson ${\bar D} D$ 
(${\bar D}^0 D^0$ and $D^- D^+$) production cross sections in these 
reactions, is important because the charmonium states above the open charm 
threshold will generally be identified by means of their decays to ${\bar D} 
D$ channels if allowed~\cite{wie11,pre15}.
 
Studies of the production and spectroscopy of charm-baryons (e.g. 
$\Lambda_c^+$) are similarly interesting. In contrast to the mesons, there 
can be more states of these systems as there are more possibilities of 
orbital excitations due the presence of three quarks. At higher ${\bar p}$ 
beam momenta at the ${\bar P}ANDA$ facility the yields of the channels with 
charm-baryons exceed those of the charm-meson channels by factors of 3-4, 
which is confirmed by calculations of the productions of ${\bar\Lambda}_c^- 
\Lambda_c^+$ and ${\bar D} D$ pairs in the ${\bar p} p$ collisions in Refs.
\cite{shy14,shy16b,hai10,hai14}. In studies of the charm-baryon production 
in the ${\bar p}$ induced reactions on proton or nuclei the production of 
extra particles is not needed for the charm conservation, which reduces 
the threshold energy as compared to, say, $pp$ collisions. Investigations  
of the charm-baryon (and also charm-meson) productions in the 
${\bar p}$-nucleus collisions explore the properties of charm-hadrons 
in the nuclear medium and provide information about the charm 
hadron-nucleon ($N$) interaction in the nuclear medium
\cite{shy16a,hos16,yam16}.

The $\Lambda_c^+ - N$ interaction has come in focus after discoveries of 
many exotic hadrons [e.g, $X(3872)$, and $Z(4430)$] by the Belle 
experiments~\cite{cho03,cho08}. These hadrons are considered to be either 
the 4-quark bound states including the charm one or the composite states 
of two (or more) hadrons (see, e.g., Ref.~\cite{god08} for a review). 
There is no conclusive evidence for the existence of the two-body bound 
states in the $\Lambda_c^+ - N$ channel. It will critically depend on the 
nature of the $\Lambda_c^+ - N$ interaction. Because performing scattering 
experiments in this channel is not feasible for the time being, 
alternative methods will have to be explored for determining this 
interaction. Some effort has been made in this direction in the lattice 
QCD calculations by the HAL QCD collaboration~\cite{miy16}, but their 
results are limited to pion masses around or in excess of 600 MeV.  
Another viable alternative is to study $\Lambda_c^+$ hypernuclei that can 
be produced in the ${\bar p}$ induced reactions on nuclei at the 
${\bar P}ANDA$ facility. In the past the study of the $\Lambda$ 
hypernuclear states has provided important information about the 
$\Lambda - N$ interaction (see, e.g., the reviews
\cite{has06,shy08,gal16}). Furthermore, in a theoretical study of the 
$\Xi$-hypernuclei, it has been shown that the properties of the states 
of such nuclei are strongly dependent on the nature of the $\Xi - N$ 
interaction~\cite{shy12}.    
  
The existence of the $\Lambda_c^+$ hypernuclei was predicted already 
in 1975~\cite{tya75}. Since then several theoretical calculations have 
been reported for such nuclei that are based on the idea of the close 
similarity between the quark structures of $\Lambda$ and $\Lambda_c^+$. 
They have predicted a rich spectrum of charm-hypernuclei spanning over 
a wide range of atomic mass numbers
\cite{iwa77,dov77,gat78,kol81,bha81,ban82,gib83,cai03}. However, there 
is a large variation in the binding energies and the potential depths 
predicted by these authors. More recently, in a series of publications
\cite{tsu03a,tsu03b,tsu03c,tsu04} systematic and quantitative studies 
of the $\Lambda_c^+$ hypernuclei have been reported within the quark-meson 
coupling (QMC) model. These studies have predicted a number of states and 
their binding energies for the $\Lambda_c^+$ hypernuclei, 
$^{17}_{\Lambda_c^+}$O, $^{41}_{\Lambda_c^+}$Ca, 
$^{49}_{\Lambda_c^+}$Ca, $^{91}_{\Lambda_c^+}$Zr, and 
$^{209}_{\Lambda_c^+}$Pb. However, none of these references reports 
any production cross section for the formation of the $\Lambda_c^+$ 
hypernuclei in an actual reaction.

In this letter, we present, for the first time, results for the cross 
sections of the $\Lambda_c^+$ hypernuclear production in the ${\bar p}$ 
- $^{16}$O collisions. We describe this reaction within an effective 
Lagrangian model (see, e.g. Refs.~\cite{shy12,shy04,shy06,shy09,ben10}), 
where ${\bar \Lambda}_c^- \Lambda_c^+$ production takes place via 
$t$-channel exchanges of $D^0$ and $D^{*0}$ mesons in collisions of the 
${\bar p}$ with one of the protons of the target nucleus in the initial 
state [see, Figs.~1(a) and 1(b)]. The $\Lambda_c^+$ is captured into 
one of the orbits of the residual nucleus to make the hypernucleus, 
while ${\bar \Lambda}_c^-$ rescatters onto its mass shell [Fig. 1(a)].  
The $s$- and $u$-channel resonance excitation diagrams are suppressed, 
as no resonance is known with energy in excess of 3.0 GeV having 
branching ratios for decay to the $\Lambda_c^+$ channel. The direct 
${\bar p}p$ annihilation into ${\bar \Lambda}_c^- \Lambda_c^+ $ via 
the contact diagrams is also suppressed due to the Okubo-Zweig-Iizuka 
rule.

It may be noted that the free space ${\bar\Lambda}_c^- \Lambda_c^+$
productions in the ${\bar p} p$ collisions have been studied in Refs.
\cite{hai10,hai17} within the J\"ulich meson-baryon model in a 
coupled-channel approach. In these studies, where calculations are 
confined to antiproton beam momenta lying close to the charm baryon 
production threshold, the coupled-channel effects have been found 
to be very important.  Even though the role of the coupled-channel
effects has not been studied at higher ${\bar p}$ beam momenta 
of relevance to the ${\bar P}ANDA$ experiment, which are of interest 
to our work, these effects may not be as strong at such higher beam 
momenta as they are at near threshold beam momenta. Inclusion of 
the coupled-channel effects of the kind discussed in Refs.
\cite{hai10,hai17} is currently out of the scope of our effective 
Lagrangian model.

Our model retains the full structure of the interaction vertices and 
treats baryons as Dirac particles. The $\Lambda_c^+$ bound state 
spinors have been calculated within the QMC model. In this model
\cite{gui88}, quarks within the non-overlapping nucleon bags (modeled 
using the MIT bag), interact self-consistently with the 
isoscalar-scalar ($\sigma$) and isoscalar-vector ($\omega$) mesons in 
the mean field approximation. The explicit treatment of the nucleon 
internal structure represents an important departure from quantum 
hadrodynamics (QHD)~\cite{ser86}. The self-consistent response of bound 
quarks to the mean $\sigma$ field leads to a new saturation mechanism 
for nuclear matter~\cite{gui88}. The QMC model has been used to study 
the properties of finite nuclei~\cite{sai96}, the binding of $\omega$, 
$\eta$, $\eta^\prime$ and $D$ mesic nuclei~\cite{tsu98a,tsu98b,tsu99} 
and also the effect of the medium on $K^\pm$ and $J/\Psi$ production
\cite{sai07}.

\begin{figure}[!t]
\begin{center}
\includegraphics[scale=.50]{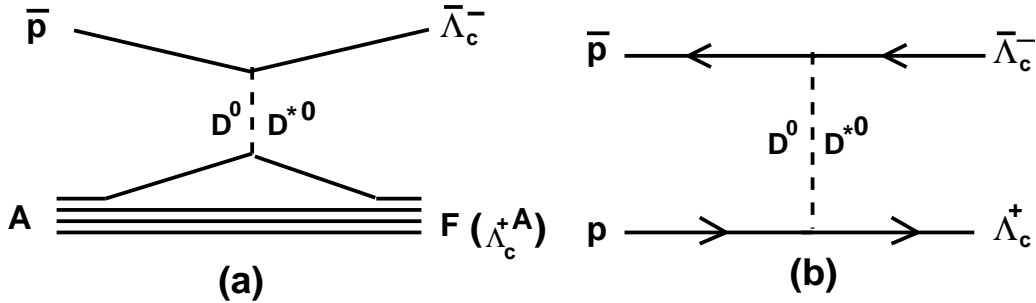}
\end{center}
\caption{
(a) Graphical representation of the model used to describe the 
charm-hypernuclear production reaction ${\bar p} + A \to {\bar \Lambda}_c^- 
+ F ( = {_{\Lambda_c^+}}A)$. $D^0$ and $D^{*0}$ in the intermediate line 
represent the exchanges of $D^0$ pseudoscalar and $D^{*0}$ vector mesons, 
respectively. (b) The diagram to describe the elementary reaction 
${\bar p} + p \to {\bar \Lambda}_c^- + \Lambda_c^+$. The arrows show the 
directions of the relative momenta.
}
\label{fig:Fig1}
\end{figure}

For the evaluation of the amplitudes corresponding to the processes shown 
in Fig.~1(a), one requires the effective Lagrangians at the 
baryon-meson-nucleon vertices, and the propagators for the $D^0$ and $D^{*0}$ 
mesons. The masses of these mesons have been taken to be 1.867 GeV and  2.008 
GeV, respectively. The denominators of these propagators involve the meson 
self-energies that account for the medium effects on their propagation through 
the nucleus. 
  
The effective Lagrangians at the charm-baryon-meson-nucleon vertices have been 
adopted from Refs.~\cite{shy14,gro90,hai11a,hob14} . For the $D^0$ 
meson-exchange vertices we have
\begin{eqnarray}
{\cal L}_{D^0BN} & = & ig_{BD^0N}{\bar \psi}_B \gamma_5 \psi_N \phi_{D^0} + 
H.c.,
\label{eq.1}
\end{eqnarray}
where  ${\psi}_B$ and $\psi_N$ are the charm-baryon and nucleon (antinucleon) 
fields, respectively, while $\phi_{D^0}$ is the $D^0$ meson field. 
$g_{BD^0N}$ in Eq.~(1) represents the vertex coupling constant at 
the charm-baryon(B)-$D^0$-meson-nucleon (-antinucleon) vertex.  

For the $D^{*0}$ meson-exchange vertices, the effective Lagrangian is
\begin{eqnarray}
{\cal L}_{D^{*0}BN} & = & g_{D^{*0}BN} {\bar \psi}_B \gamma_\mu \psi_N 
                      \theta_{D^{*0}}^\mu + \frac{f_{D^{*0}BN}}{4M}
                     {\bar \psi}_B \sigma_{\mu \nu} \psi_N 
                     G_{D^{*0}}^{\mu \nu} + H.c.,
\label{eq.2}
\end{eqnarray}
where $\theta_{D^{*0}}^\mu$ is the vector meson field, with field strength 
tensor $G_{D^{*0}}^{\mu \nu} = \partial^\mu \theta_{D^{*0}}^\nu - \partial^\nu 
\theta_{D^{*0}}^\mu$, and $\sigma_{\mu \nu}$ is the usual tensor operator in 
the Dirac space. The vector and tensor couplings are defined by $g$ and $f$, 
respectively.  

The values of $g$ and $f$ at various vertices are adopted from Refs.
\cite{hai11a,hai07,hai08}, as $g_{ND^0B}$ = 13.98, $g_{ND^{*0}B}$ = 5.64 and 
$f_{ND^{*0}B}$ = 18.37. We add that the same coupling constants were used in 
the description of the free-space charmed hadron production in ${\bar p} p $ 
collisions within the effective Lagrangian model~\cite{shy14,shy16b} as well 
as the J\"ulich meson-exchange model~\cite{hai10,hai14,hai17}. Furthermore,
same values of the vector and tensor couplings were also used in Ref.
\cite{hob14} in investigations of the role of intrinsic charm in the nucleon 
using a model formulated in terms of effective meson-baryon degrees of 
freedom.   

The off-shell corrections at various vertices are accounted for by 
introducing form factors $(F_i)$. In our study of the free-space ${\bar 
\Lambda}_c^- + \Lambda_c^+$ production~\cite{shy14}, monopole form 
factors~\cite{shy99,shy02} were used at all the vertices. In order to 
maintain consistency with these calculations, we employ form factors of 
the same shape (monopole) to regulate the off-shell behavior of the 
vertices in the present work.  We write 
\begin{eqnarray}
F_i(q_{D_i}) & = & \frac{\lambda_i^2-m_{D_i}^2}{\lambda_i^2-q_{D_i}^2},
\label{eq.3}
\end{eqnarray}
with a cutoff parameter ($\lambda_i$) of 3.0 GeV, which is the same as that 
used in Ref.~\cite{shy14}. In Eq.~{\ref{eq.3}, $q_{D_i}$ is the momentum of 
the {\it i-}th exchanged meson with mass $m_{D_i}$. It may be noted that the 
same shape of the form factor with the same $\lambda_i$ was also used in the 
studies of the ${\bar D} D$ production in ${\bar p} p$ and ${\bar p}$-nucleus 
collisions in Refs.~\cite{shy16b} and \cite{shy16a}, respectively, within a 
similar type of the effective Lagrangian model.  

The propagators for the $D^0$ and $D^{*0}$ mesons are given by
\begin{eqnarray}
G_{D^0}(q_{D^0}) & = & \frac{i} {{q_{D^0}^2 - m_{D^0}^2 }-\Pi_{D^0}},
\label{eq.4}\\
G_{D^{*0}}^{\mu\nu}(q_{D^{*0}}) & = & -i\left(\frac{{g^{\mu\nu}-q_{D^{*0}}^\mu 
                                  q_{D^{*0}}^\nu/q_{D^{*0}}^2}}
                           {{q_{D^{*0}}^2 - (m_{D^{*0}}-i\Gamma_{D^{*0}}/2)^2}-
                           \Pi_{D^{*0}}} \right),
\label{eq.5}
\end{eqnarray}
where $q_{D^0}$ and $q_{D^{*0}}$ are the four-momenta of $D^0$ and $D^{*0}$
mesons, respectively, while $m_{D^0}$ and $m_{D^{*0}}$ are their masses. 
$\Pi_{D^0}$ and $\Pi_{D^{*0}}$ represent the (complex) self-energies of 
$D^0$ and $D^{*0}$ mesons, respectively. In Eq.~(\ref{eq.5}), 
$\Gamma_{D^{*0}}$ is the total width of the $D^{*0}$ meson, which is about 
2.0 MeV according to the latest particle data group estimate~\cite{oli14}.

In this exploratory study, the self-energies $\Pi_{D^0}$ and $\Pi_{D^{*0}}$ 
have been obtained from the mean-field potentials for $D^0$ and $D^{*0}$
mesons in $^{16}$O calculated self-consistently within the QMC model in the 
local density approximation as described in Ref.~\cite{tsu99}. The 
self-energy is related to the potential by $\Pi_{D_i} = 
2\omega_{D_i} U_{D_i}(q_{D_i})$, where $\omega_{D_i} = 
\sqrt{q_{D_i}^2 + m_{D_i}^2}$, and $U_{D_i}$ is the potential in the 
momentum space of meson $D_i$, with $D_i$ standing for $D^0$ or $D^{*0}$.

After having established the effective Lagrangians, coupling constants, and
forms of the propagators, the amplitudes of $D^0$ and $D^{*0}$ exchange 
diagrams can be easily written. The signs of these amplitudes are fixed by 
those of the effective Lagrangians, the coupling constants, and the 
propagators as described above.  These signs are not allowed to change 
anywhere in the calculations. 

At the upper vertices of Fig.~1(a), the amplitudes involve free-space 
spinors of the antiparticles, while at the lower vertices, they have 
spinors for bound proton in the initial state (to be represented by 
$\psi(k_p)$) and bound $\Lambda_c^+$ in the final state (to be represented 
by $\psi(k_{\Lambda_c^+}))$. These are the four component Dirac spinors, 
which are solutions of the Dirac equation for a bound state problem in the 
presence of external potential fields. They are calculated within the QMC 
model as described below. We write  

\begin{eqnarray}
\psi(k_i) & = & \delta(k_{i0}-E_i)\begin{pmatrix}
                    {f(K_i) {\cal Y}_{\ell 1/2 j}^{m_j} (\hat {k}_i)}\\
                    {-ig(K_i){\cal Y}_{\ell^\prime 1/2 j}^{m_j}
                     (\hat {k}_i)} 
                    \end{pmatrix}. \label{eq.6}
\end{eqnarray}

In our notation $k_i$ represents a four momentum, and $\vec{k_i}$ a three
momentum. The magnitude of $\vec {k_i}$ is represented by $K_i$, and its
directions by $\hat {k}_i$. $k_{i0}$ represents the timelike component of
momentum $k_i$. In Eq.~(\ref{eq.6}), $f(K_i)$ and $g(K_i)$ are the radial 
parts of the upper and lower components of the spinor $\psi(k_i)$ with $i$ 
representing either a proton or a $\Lambda_c^+$. The coupled spherical 
harmonics, ${\cal Y}_{\ell 1/2 j}^{m_j}$,  is given by
\begin{eqnarray}
{\cal Y}_{\ell 1/2 j}^{m_j}(\hat {k}_i) & = & <\ell m_\ell 1/2 \mu | j m_j>
                        Y_{\ell m_\ell}(\hat {k}_i) \chi_{\mu},\label{eq.7}
\end{eqnarray}
where $Y_{\ell m_\ell}$ represents the spherical harmonics, and $\chi_{\mu}$ the
spin-space wave function of a spin-$\frac{1}{2}$ particle. In Eq.~(\ref{eq.6})
$\ell^\prime = 2j - \ell$ with $\ell$ and $j$ being the orbital and total
angular momenta, respectively.
  
To calculate the spinors for the final bound charm-hypernuclear state and 
the initial proton bound state within the QMC model, we construct a simple, 
relativistic shell model, with the nucleon core calculated in a combination 
of self-consistent scalar and vector mean fields. The Lagrangian density 
for a hypernuclear system in the QMC model is written as a sum of two terms, 
\begin{eqnarray}
{\cal L}^{HY}_{QMC} &=& {\cal L}^N_{QMC} + {\cal L}^Y_{QMC},
\label{eq:LagYQMC} \\
{\cal L}^N_{QMC} &\equiv&  \overline{\psi}_N(\vec{r})
\left[ i \gamma \cdot \partial
- M_N^*(\sigma) - (\, g_\omega \omega(\vec{r})
+ g_\rho \frac{\tau^N_3}{2} b(\vec{r})
+ \frac{e}{2} (1+\tau^N_3) A(\vec{r}) \,) \gamma_0
\right] \psi_N(\vec{r}) \quad \nn \\
  & & - \frac{1}{2}[ (\nabla \sigma(\vec{r}))^2 +
m_{\sigma}^2 \sigma(\vec{r})^2 ]
+ \frac{1}{2}[ (\nabla \omega(\vec{r}))^2 + m_{\omega}^2
\omega(\vec{r})^2 ] \nn \\
 & & + \frac{1}{2}[ (\nabla b(\vec{r}))^2 + m_{\rho}^2 b(\vec{r})^2 ]
+ \frac{1}{2} (\nabla A(\vec{r}))^2, \label{eq:LagN} \\
{\cal L}^Y_{QMC} &\equiv&
\overline{\psi}_Y(\vec{r})
\left[ i \gamma \cdot \partial
- M_Y^*(\sigma)
- (\, g^Y_\omega \omega(\vec{r})
+ g^Y_\rho I^Y_3 b(\vec{r})
+ e Q_Y A(\vec{r}) \,) \gamma_0
\right] \psi_Y(\vec{r}), 
\nn\\
& & (Y = \Lambda,\Sigma^{0,\pm},\Xi^{0,+},
\Lambda^+_c,\Sigma_c^{0,+,++},\Xi_c^{0,+},\Lambda_b),
\label{eq:LagY}
\end{eqnarray}
where $\psi_N(\vec{r})$ and $\psi_Y(\vec{r})$ are the nucleon and the
hyperon (strange, charm or bottom baryon) fields, respectively.
$A(r)$ is the Coulomb field.
$g_\omega$ and $g_{\rho}$ are the $\omega$-N and $\rho$-N coupling constants
which are related to the corresponding $(u,d)$ quark-$\omega$, $g_\omega^q$,
and $(u,d)$ quark-$\rho$, $g_\rho^q$, coupling constants, as $g_\omega = 3 
g_\omega^q$ and $g_\rho = g_\rho^q$.  $I^Y_3$ and $Q_Y$ are the third
component of the hyperon isospin operator and its electric charge in units
of the proton charge, $e$, respectively.

The following set of equations of motion are obtained for the hypernuclear
system from the Lagrangian densities defined by 
Eqs.~(\ref{eq:LagN})-(\ref{eq:LagY}),
\begin{eqnarray}
&&\hspace{-1.4cm}[i\gamma \cdot \partial -M_N(\sigma)-
(\, g_\omega \omega(\vec{r}) + g_\rho \frac{\tau^N_3}{2} b(\vec{r})  
+ \frac{e}{2} (1+\tau^N_3) A(\vec{r}) \,) \gamma_0 ] \psi_N(\vec{r})
=  0, 
\label{eqdiracn1}\\
%%%%%
&&\hspace{-1.4cm}[i\gamma \cdot \partial - M_Y(\sigma)-
(\, g^Y_\omega \omega(\vec{r}) + g_\rho I^Y_3 b(\vec{r}) 
 + e Q_Y A(\vec{r}) \,) \gamma_0 ] \psi_Y(\vec{r}) = 0, 
\label{eqdiracy2}\\
%%%%%
&&(-\nabla^2_r+m^2_\sigma)\sigma(\vec{r})  = 
g_\sigma C_N(\sigma) \rho_s(\vec{r})
 + g^Y_\sigma C_Y(\sigma) \rho^Y_s(\vec{r}), 
\label{eqsigma}\\
%%%%%
&&\hspace{-1.4cm}(-\nabla^2_r+m^2_\omega) \omega(\vec{r})  =
g_\omega \rho_B(\vec{r}) + g^Y_\omega
\rho^Y_B(\vec{r}),
\label{eqomega}\\
%%%%%%
&&\hspace{-1.4cm}(-\nabla^2_r+m^2_\rho) b(\vec{r})  =
\frac{g_\rho}{2}\rho_3(\vec{r}) + g^Y_\rho I^Y_3 \rho^Y_B(\vec{r}),
\label{eqrho}\\
%%%%%%
&&\hspace{-1.4cm}(-\nabla^2_r) A(\vec{r})  =
e \rho_p(\vec{r})
+ e Q_Y \rho^Y_B(\vec{r}) ,
\label{eqcoulomb}
\end{eqnarray}
where, $\rho_s(\vec{r})$ ($\rho^Y_s(\vec{r})$), $\rho_B(\vec{r})$
($\rho^Y_B(\vec{r})$), $\rho_3(\vec{r})$ and $\rho_p(\vec{r})$ are the
scalar, baryon, third component of isovector, and proton densities at 
position $\vec{r}$ in the hypernucleus~\cite{tsu98a}. On the right hand 
side of Eq.~(\ref{eqsigma}), a new and characteristic feature of the QMC 
model appears that arises from the internal structures of the nucleon 
and hyperon, namely, $g_\sigma C_N(\sigma)= - \frac{\partial M_N(\sigma)}
{\partial \sigma}$ and $g^Y_\sigma C_Y(\sigma)= - \frac{\partial 
M_Y(\sigma)}{\partial \sigma}$ where $g_\sigma \equiv g_\sigma (\sigma=0)$ 
and $g^Y_\sigma \equiv g^Y_\sigma (\sigma=0)$. We use the nucleon and 
hyperon masses to calculate $C_{N,Y}(\sigma)$ employing the parameters of QMC-I 
summarized in table 13 of Ref.~\cite{sai07}. The scalar and vector fields 
as well as the spinors for hyperons and nucleons, can be obtained by 
solving these coupled equations self-consistently.
\begin{table}
\begin{center}
\caption{Binding energies of $\Lambda_c^+$ and $p$ for each shell as predicted 
by the QMC model.}
\vspace{0.5cm}
\begin{tabular}{ccccc}
\hline
 State & BE \\
       &(\footnotesize{MeV})\\
\hline
$^{16}\hspace{-0.3cm}{_{\Lambda_c^+}}$O($\Lambda_c^+$ 0$p_{1/2}$) & 7.17 \\
$^{16}\hspace{-0.3cm}{_{\Lambda_c^+}}$O($\Lambda_c^+$ 0$p_{3/2}$) & 7.20 \\
$^{16}\hspace{-0.3cm}{_{\Lambda_c^+}}$O($\Lambda_c^+$ 0$s_{1/2}$) & 12.78 \\
\hline
$^{16}$O(p 0$p_{1/2}$)                                & 11.87  \\
\hline
\end{tabular}
\end{center}
\label{tab_bound}
\end{table}

We note here that, for the Dirac equation for $\Lambda_c^+$ baryon 
[Eq.~(\ref{eqdiracy2})], the effects of Pauli blocking at the quark level, 
is introduced by adding a repulsive potential. This is the same as that 
used for the $\Lambda$-hyperon case. This was extracted by the fit to the 
$\Lambda$- and $\Sigma$-hypernuclei taking into account the $\Sigma N 
- \Lambda N$ channel coupling~\cite{tsu03b,tsu03c,tsu04,tsu98b}. The 
modified Dirac equation for the $\Lambda_c^+$ baryon is, 
\begin{equation}
 [i\gamma \cdot \partial - M_Y(\sigma)-
(\, \lambda_{\Lambda_c^+}\rho_B(\vec{r}) + g^Y_\omega \omega(\vec{r}) 
+ g_\rho I^Y_3 b(\vec{r}) 
 + e Q_Y A(\vec{r}) \,) \gamma_0 ] \psi_Y(\vec{r}) = 0,
\label{Pauli}
\end{equation}
where $\rho_B(\vec{r})$ is the baryon density at the position $\vec{r}$ in 
the $\Lambda_c^+$-hypernucleus. The value of $\lambda_{\Lambda_c^+}$ is 
60.25 MeV (fm)$^3$. The details about the effective Pauli blocking at 
the quark level can be found  in Refs.~\cite{tsu98b,sai07}.

\begin{figure}[!t]
\begin{center}
\includegraphics[scale=.50]{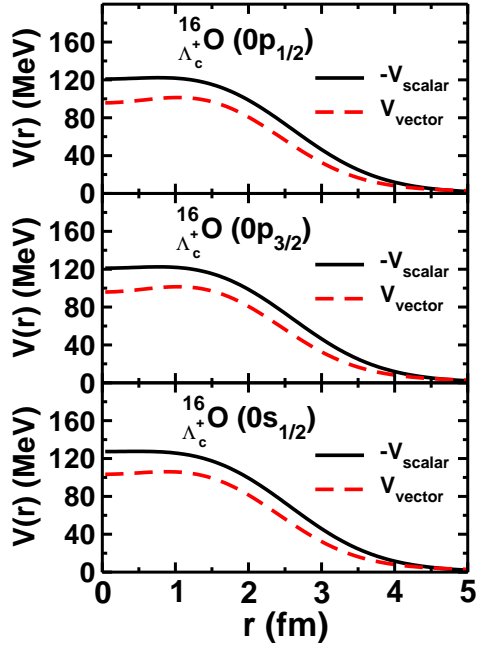}
\end{center}
\caption{(color online)
Vector and scalar potential fields felt by $\Lambda_c^+$ for 0p$_{1/2}$, 
0p$_{3/2}$ and 0s$_{1/2}$ $\Lambda_c^+$ states in $^{16}_{\Lambda_c^+}$O, 
where the vector potential field strength contains the effective Pauli 
blocking potential for each state of $\Lambda_c^+$ [see Eq.~(\ref{Pauli})].
}
\label{Fig2}
\end{figure}

We have chosen the reaction ${\bar p}$ + $^{16}$O $\to {\bar \Lambda}_c^- 
+ {^{16}_{\Lambda_c^+}}$O for the first numerical application of our 
model of charm-hypernuclear production. The target nucleus is a doubly 
closed system. The QMC model predicts three bound states for the 
charm-hypernucleus $^{16}_{\Lambda_c^+}$O. The predicted quantum numbers 
and binding energies of these states are shown in Table 1. We assume the 
initial bound proton state to have quantum numbers of the outermost 
$0p_{1/2}$ proton orbit of the target nucleus. The predicted 
binding energy of this state by QMC is also shown in Table 1. The 
hypernuclear spectrum is clearly divided into three groups corresponding 
to the configurations ($0p_{1/2}^{-p},0s_{1/2}^{\Lambda_c^+}$), 
($0p_{1/2}^{-p},0p_{1/2}^{\Lambda_c^+}$), and 
($0p_{1/2}^{-p},0p_{3/2}^{\Lambda_c^+}$). Since $0p_{3/2}$ proton hole 
state has a much larger binding energy, any configuration mixing is 
expected to be negligible and has not been considered in this study.

In Fig.~\ref{Fig2}, we show the scalar and vector fields as calculated 
within the QMC model for $0p_{1/2}$, $0p_{3/2}$ and $0s_{1/2}$ 
$\Lambda_c^+$ states, where the strength of the vector field contains also 
the effective Pauli blocking potential for each state of $\Lambda_c^+$ (see 
Eq.~(\ref{Pauli})). It may be recalled that in the QMC model the scalar and 
vector fields are generated by the couplings of the $\sigma$ and $\omega$ 
mesons to the quarks.  Due to the different masses of these mesons and 
their couplings, especially the density dependence or non-linear dependence 
of the $\sigma N$ and $\sigma {\Lambda_c^+}$ coupling strengths due to the 
baryon internal structure, the scalar and vector fields may acquire 
non-trivial radial dependence.  This is in contrast to a phenomenological 
model where scalar and vector fields have the same Woods-Saxon (WS) radial 
form (see, e.g., Ref.~\cite{shy12}).  In any case, because in such models 
the depths of these fields are searched to reproduce the experimental 
binding energies of a given state, they can not be applied at this stage 
to $\Lambda_c^+$ bound states due lack of any experimental information about 
them. In Fig.~\ref{Fig2}, we note that sum of the scalar and vectors fields 
at $r = 0$ is about -30 MeV for all the three states. This is roughly 
equivalent to the depth of a non-relativistic potential at this point. This 
is about 15 MeV less (in absolute value) than the depth of the $\Lambda_c^+$ 
- $^{16}$O Hartree potential calculated in Ref.~\cite{dov77}. Therefore, 
relativistic self-consistent procedure has its consequences. 

Figs.~3(a) and 3(b) show the moduli of the upper and lower components of 
0p$_{1/2}$, 0p$_{3/2}$ and 0s$_{1/2}$ $\Lambda_c^+$ spinors for the 
$^{16}_{\Lambda_c^+}$O charm-hypernucleus in coordinate space and momentum 
space, respectively. The spinors in the momentum space are obtained by 
Fourier transformation of the corresponding coordinate space spinors. 
We note that only for $K_{\Lambda_c^+} < $ approximately 1.0 fm$^{-1}$, 
are the magnitudes of the lower components ($|g(K_{\Lambda_c^+})|$) 
substantially smaller than those of the upper components 
($|f(K_{\Lambda_c^+})|$). In the region of $K_{\Lambda_c^+}$ pertinent 
to the charm-hypernuclear production, $|g(K_{\Lambda_c^+})|$ may not be 
negligible.

\begin{figure}[!t]
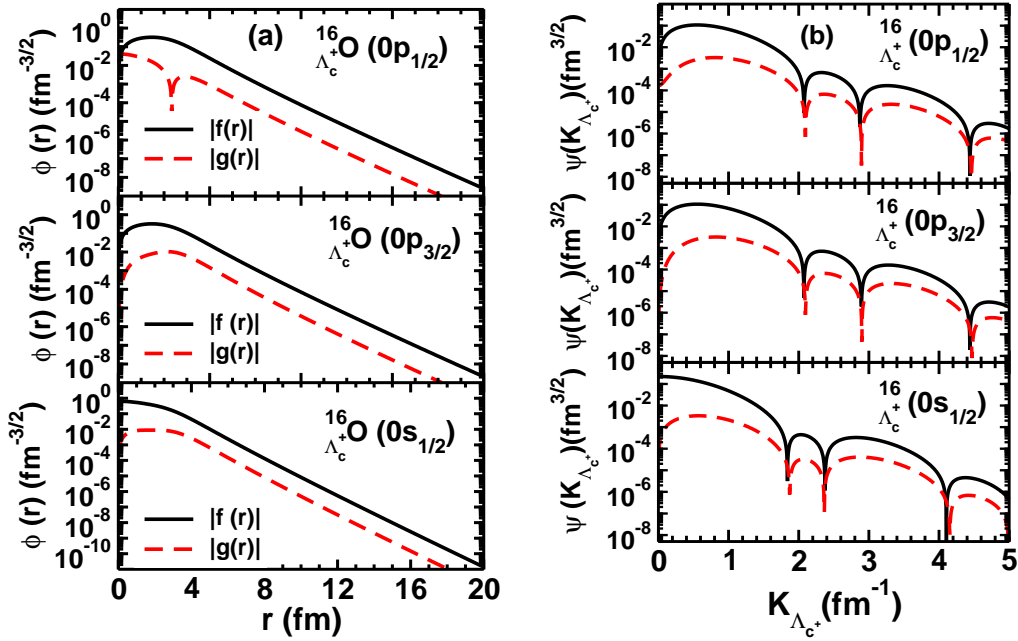

\begin{tabular}{cc}
\includegraphics[scale=.50]{charm-hyper_Fig3a.eps} & \hspace {0.05cm}
\includegraphics[scale=.48]{charm-hyper_Fig3b.eps} 
\end{tabular}
\caption{(color online)
(a) Moduli of the upper ($|f(r)|$) and lower ($|g(r)|$) components of the 
coordinate space  spinors for 0p$_{1/2}$, 0p$_{3/2}$ and 0s$_{1/2}$ 
$\Lambda_c^+$ states in  $^{16}_{\Lambda_c^+}$O. Solid lines represent the 
upper component while the dashed line the lower component. (b) Moduli of 
upper (solid lines)  and lower (dashed lines) components of the momentum 
space spinors of the $\Lambda_c^+$ bound states in  $^{16}_{\Lambda_c^+}$O 
for the same states as in the left panel.
}
\label{fig:Fig3}
\end{figure}

To get the $T$-matrix of the reaction of the process shown in Fig.~1(a), 
one has to integrate the amplitudes corresponding to each meson-exchange 
graph over the independent intermediate momenta $k_p$ and $k_{\Lambda_c^+}$ 
subject to the constraints imposed by the momentum conservation at each 
vertex.

We have used plane waves to describe the motions of the ${\bar p}$ and
${\bar \Lambda}_c^-$ baryon in the entrance and outgoing channels, 
respectively.  However, initial- and final-state interactions are 
approximately accounted for within an eikonal-approximation based 
procedure~\cite{shy06} that was used in  Refs.~\cite{shy16a} to describe 
the charm-meson production in the ${\bar p}$-nucleus reactions. In this 
work we have employed the same parameters as those of Ref.~\cite{shy16a} 
to estimate the distortion effects in the initial and final channels.

The threshold momentum for the production of $^{16}_{\Lambda_c^+}$O in 
${\bar p}$ induced reaction on $^{16}$O is about 3.953 GeV/c, while that
for the ${\bar \Lambda}_c^- + \Lambda_c^+$ production in the elementary  
${\bar p} + p$ reaction is 10.162 GeV/c. The shift in the threshold of
the hypernuclear production reaction to lower beam momenta is mainly due 
to the Fermi motion effects. Therefore, $\Lambda_c^+$ hypernuclear 
production experiments may be feasible even in the beginning stage of 
the FAIR project when ${\bar p}$ beam momenta may be lower than their 
maximum planned value of 15 GeV/c. 

In Fig.~4, we show the $0^\circ$ differential cross sections 
$[(d\sigma/d\Omega)_0]$ for the reaction 
${\bar p}$ + $^{16}$O $\to {\bar \Lambda}_c^- + {^{16}_{\Lambda_c^+}}$O 
obtained by using the proton-hole and $\Lambda_c^+$ bound state spinors 
calculated within the QMC model. Cross sections are shown for ${\bar p}$ 
beam momenta in the range of threshold to 20 GeV/c. The 
charm-hypernuclear states populated are $1^-$ and $0^-$, $1^+$ and 
$0^+$, and $2^+$ and $1^+$ corresponding to the particle-hole 
configurations ($0p_{1/2}^{-p},0s_{1/2}^{\Lambda_c^+}$),
($0p_{1/2}^{-p},0p_{1/2}^{\Lambda_c^+}$), and
($0p_{1/2}^{-p},0p_{3/2}^{\Lambda_c^+}$), respectively. Cross sections 
to the higher $J$ state of each configuration are shown in Fig.~4(a), 
while those to lower $J$ in Fig.~4(b).  We see that for each 
particle-hole configuration, the state with higher $J$ has larger cross 
section. While for beam momenta $\ge $ 8 GeV/c, the cross-sections for 
states belonging to the ($0p_{1/2}^{-p},0s_{1/2}^{\Lambda_c^+}$) 
configuration are larger than those of the other two configurations, 
for $p_{\bar p}$ lower than 8 GeV/c those for states belonging to the 
($0p_{1/2}^{-p},0p_{1/2}^{\Lambda_c^+}$) configuration have larger 
values. However, the maximum difference between the cross sections of 
the states belonging to two configurations is not more than a factor of 
1.4. The  $[(d\sigma/d\Omega)_0]$, for the states belonging to the 
configuration  ($0p_{1/2}^{-p},0p_{3/2}^{\Lambda_c^+}$) are smaller 
than those of the ($0p_{1/2}^{-p},0p_{1/2}^{\Lambda_c^+}$) 
configuration by factors of approximately 1.4 - 3.0. 

\begin{figure}[!t]
\begin{center}
\includegraphics[scale=.60]{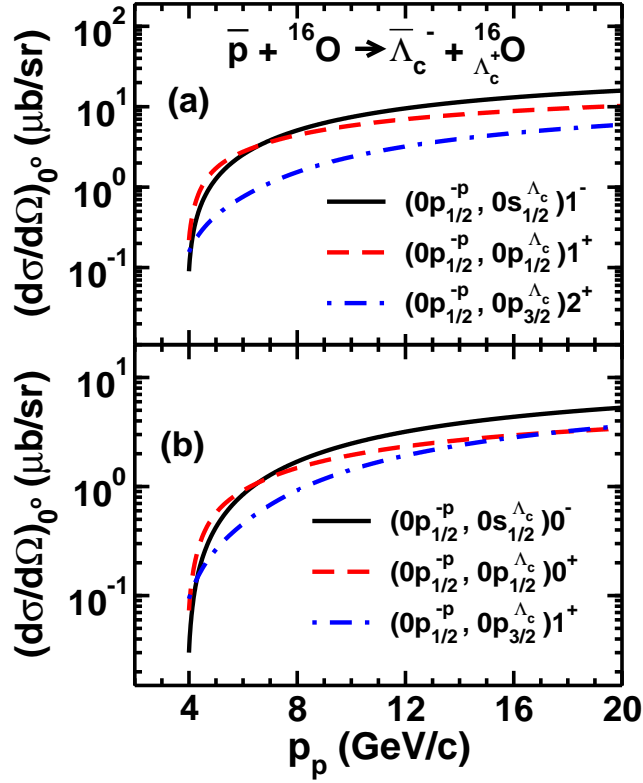} 
\end{center}
\caption{(color online)
(a) Differential cross sections at $0^\circ$ of the ${\bar p}$ + $^{16}$O
$\to {\bar \Lambda}_c^- + {^{16}_{\Lambda_c^+}}$O reaction leading to the
$^{16}_{\Lambda_c^+}$O hypernuclear states of larger $J$ value of each 
particle-hole configuration as indicated. (b) The same as in (a) but for
states of lower $J$ value of each configuration as indicated. In the 
legends $\Lambda_c$ corresponds to $\Lambda_c^+$. 
}
\label{fig:Fig4}
\end{figure}

It is to be noted that for ${\bar p}$ beam momenta between 8 - 15 GeV/c 
(which are of interest to ${\bar P}ANDA$ experiment) the magnitudes of the 
$0^\circ$ differential cross sections vary between 1.5 - 3.8 $\mu$b/sr,
and 5.0 - 11.0 $\mu$b/sr for states $0^-$ and $1^-$, respectively, of the 
configuration ($0p_{1/2}^{-p},0s_{1/2}^{\Lambda_c^+}$). On the other hand, 
for states $1^+$ and $2^+$ of the configuration 
($0p_{1/2}^{-p},0p_{3/2}^{\Lambda_c^+}$), it varies between 0.9 - 2.8 
$\mu$b/sr, and 1.6 - 6.0 $\mu$b/sr, respectively, Furthermore, the cross 
sections for all the states consist almost solely of the contributions from 
the $D^{*0}$ meson-exchange process, with $D^0$ meson-exchange terms being 
very small. This is similar to what has been noted in the case of the 
elementary ${\bar p} + p \to {\bar \Lambda}_c^- + \Lambda_c^+$ reaction in 
Ref.~\cite{shy14}. The domination of the $D^{*0}$ exchange process has been 
traced to the strong tensor coupling term of the $D^{*0}$ meson couplings. 
 
We next discuss the uncertainties and the range of validity of our results.
We use an eikonal-approximation based phenomenological method to account 
for the initial- and final-state interaction effects. In the recent past
\cite{yam16,lar17a}, similar methods have been used for this purpose 
in other studies also where models like ours have been employed to 
investigate the hypernuclear productions of different types in the 
${\bar p}$-nucleus collisions. However, due to lack of the adequate 
experimental information on the elastic scattering data these methods 
involve parameters, which may not be properly constrained. This 
constitutes a major source of uncertainty in such calculations.

Our procedure~\cite{shy16a} involves basically two parameters; namely, 
the ${\bar p}$-nucleon and ${\bar \Lambda}_c^-$-nucleon total cross 
sections ($\sigma_{{\bar p}N}$, and $\sigma_{{\bar \Lambda}_c^-N}$, 
respectively). Out of these $\sigma_{{\bar p}N}$ is of crucial importance. 
Its value has been quoted to be some where between 50 mb and 81 mb for 
${\bar p}$ momenta above 2 GeV/c (see, Refs.~\cite{lar17b,tec14,shu96}). 
The results shown in Figs.~4(a) and 4(b) have been obtained with a 
$\sigma_{{\bar p}N}$ of 75 mb.  Nevertheless, we have also performed 
calculations by using $\sigma_{{\bar p}N}$ of 50 mb and 81 mb in order 
to access the uncertainty in the magnitudes of our cross sections. For 
the ($0p_{1/2}^{-p},0p_{1/2}^{\Lambda_c^+}$) configuration, at the 
${\bar p}$ beam momentum of 15 GeV/c, the cross sections increase by 
factor of almost 2 or decrease by about 10$\%$ by using 
$\sigma_{{\bar p}N}$ = 50 mb and 81 mb, respectively, as compared to 
those obtained with $\sigma_{{\bar p}N}$ = 75 mb. Thus, there could to 
be an uncertainty of a factor of up to 2 in our cross sections on this 
account.  Furthermore, our method necessarily assumes that the shapes 
of the angular distributions are not influenced by the distortion 
effects. This may change in a more rigorous treatment of distortions 
that includes both absorption and dispersion effects. 

Results of the calculations performed within our model are also 
sensitively dependent on the values of the coupling constants at 
various vertices involved in the $t$-channel diagrams, and on the 
shape of the form factor and the value of the cutoff parameter 
$\lambda_i$.

As stated earlier, the coupling constants at the vertices involved in 
the $t$-channel diagrams, have been adopted from Refs.
\cite{hai11a,hai07,hai08}, where they have been fixed by using the 
SU(4) symmetry arguments in the description of the exclusive charmed 
hadron production in the ${\bar D}N$ and $DN$ scattering within a 
one-boson-exchange picture.  While, we acknowledge that the SU(4) 
symmetry will not hold rigorously, the level of deviations from the 
SU(4) coupling constants in the charm sector has been reported to be 
highly model dependent~\cite{kre12}. Recent calculations within 
light-cone sum rules suggest that deviations from the SU(4) values of 
the relevant coupling constants are limited to factors of 2 or less
\cite{kho12}. On the other hand, in  constituent quark model 
calculations using the $^3P_0$ quark-pair creation mechanism, the 
deviations are at the most of the order of 30$\%$~\cite{kre14}.

There may indeed be some uncertainty in our cross sections coming from 
the shape of the form factor and the value of the cutoff parameter 
($\lambda_i$) involved therein. We have employed a monopole form factor 
with a $\lambda_i$ of 3.0 GeV. Obviously, a form factor of a different 
shape and/or a different value of $\lambda_i$ would lead to a different 
cross section.  We have tried to minimize such uncertainties in our 
cross sections by using the same shape (monopole) of the form factor and 
the same value of $\lambda_i$ (3.0 GeV) that were used in our previous 
study of the free-space charmed baryon production in the ${\bar p} p$ 
collisions~\cite{shy14}. The same form factor with the same cutoff 
parameter were also used in the calculations of the charm baryon 
production cross sections within the J\"ulich meson-exchange model
\cite{hai10}. Moreover, this ansatz for the form factor and the value 
of the cutoff parameter $\lambda_i$ have been checked by fitting the data 
on the $pp \to \Lambda_c^+ X$ reaction measured by the ISR Collaboration 
in Ref.~\cite{hob14}. 

In summary, we have studied the $0^\circ$ differential cross sections 
for the production of the charm-hypernucleus $^{16}_{\Lambda_c^+}$O in 
the antiproton - $^{16}$O collisions within a covariant model. In our 
calculations, ${\bar \Lambda}_c^- \Lambda_c^+$ production takes place via 
the $t$-channel exchanges of $D^0$ and $D^{*0}$ mesons in the initial 
collisions of the ${\bar p}$ with a target proton. $\Lambda_c^+$ 
gets captured into one of the nuclear orbits while ${\bar \Lambda}_c^-$ 
goes out. $\Lambda_c^+$ bound state spinors are derived from the 
quark-meson coupling model. The coupling constants at the meson exchange 
vertices are taken to be the same as those used in a previous study of 
the elementary ${\bar p} + p \to {\bar \Lambda}_c^- + \Lambda_c^+$ 
reaction by one of us~\cite{shy14}. The off-shell corrections at the 
vertices are accounted for by introducing monopole form factors with a 
cut-off parameter of 3.0 GeV, which is the same as that used in 
Ref.~\cite{shy14}. 
 
At beam momenta of interest to the ${\bar P}ANDA$ experiment, the  
$0^\circ$ differential cross section for the ${\bar p}$ + $^{16}$O 
$\to {\bar \Lambda}_c^- + {^{16}_{\Lambda_c^+}}$O reaction varies 
between 0.9 $\mu$b/sr to 11 $\mu$b/sr depending on the final $\Lambda_c^+$ 
state excited in the reaction. The relatively larger cross sections and low 
threshold for the production of the $^{16}_{\Lambda_c^+}$O hypernuclear 
states in the $\bar p$ - $^{16}$O reaction may make it possible to perform 
such experiments at the ${\bar P}ANDA$ facility even in the beginning stage 
of the FAIR.

We acknowledge that there are uncertainties in our cross sections due to the 
imprecise knowledge of the parameters involved in the treatment of  initial- 
and final-state interaction effects. We provided estimates of these 
uncertainties. Accumulation of the precise relevant data by the 
${\bar P}ANDA$ experiment at the FAIR facility will help in having a 
better understanding of these effects. 
 
This work has been supported by the Science and Engineering Research Board 
(SERB), Department of Science and Technology, Government of India under 
Grant no. SB/S2/HEP-024/2013, and by Funda\c{c}\~{a}o de Amparo \`{a} 
Pesquisa do Estado de S\~{a}o Paulo (FAPESP), Brazil, Grants No.~2016/04191-3, 
and, No.~2015/17234-0, and Conselho Nacional de Desenvolvimento Cientifico e 
Tecnol\`{o}gico (CNPq), Grants No.~400826/2014-3 and No.~308088/2015-8.

\end{document}